# FEASIBLE METHODOLOGY FOR OPTIMIZATION OF A NOVEL REVERSIBLE BINARY COMPRESSOR

Neeraj Kumar Misra, Mukesh Kumar Kushwaha, Subodh Wairya and Amit Kumar

Department of Electronics Engineering,
Institute of Engineering and Technology, Lucknow, India

## ABSTRACT

*Now a day's reversible logic is an attractive research area due to its low power consumption in the area of VLSI circuit design. The reversible logic gate is utilized to optimize power consumption by a feature of retrieving input logic from an output logic because of bijective mapping between input and output. In this manuscript, we design 4:2 and 5:2 reversible compressor circuits using a new type of reversible gate. In addition, we propose new gate, named as inventive0 gate for optimizing a compressor circuit. The utility of the inventive0 gate is that it can be used as full adder and full subtraction with low value of garbage outputs and quantum cost. An algorithm is shown for designing a compressor structure. The comparative study shows that the proposed compressor structure outperforms the existing ones in terms of garbage outputs, number of gates and quantum cost. The compressor can reduce the effect of carry (Produce from full adder) of the arithmetic frame design. In addition, we implement a basic reversible gate of MOS transistor with less number of MOS transistor count.*

## 1. INTRODUCTION

In recent days a reversible design is becoming more popular due to its low power consumption. According to R.Landauer [1], the amount of heat dissipated for every bit of computation information, regardless of its synthesis is KT ln2, where k is the Boltzmann's constant $(1.380x10^{-23})$ and T is the absolute temperature. C.H. Bennett [2] proves that the heat loss in a circuit can be omitted by designing reversible structure by utilizing reversible gates and without feedback and fan-out concept. Reversible design is based on two concepts: logic reversibility means input and output must be uniquely retrievable from each other and second concept that logic work in backward and in each operation no heat dissipation and satisfies physical reversibility [5, 13, 15]. Therefore, they potentially help to solve at least two problems: overheating and power saving [7, 8, 10, 16]. The reversible logic may be especially important in low power VLSI application.

The compressor has wide application in reducing power during the frame of the partial product design and the different types of addition operations design [4, 14, 15, 21, 22]. The compressor reduces the critical path, architectural complexity and the impact of carry generated from full adder operations. The manuscript is organised with the following sections: Section 2 gives the basic definition of reversible logic and its inverse property. Section 3 shows the basic reversible gate in MOS transistor realization. Section 4 introduces new reversible compressor structure like 4:2 and 5:2 compressor circuits with it discuss parameter. Section 5 gives the comparative analysis table with its existing design. Finally, the manuscript is concluded and references.

## 2. BASIC DEFINITIONS RELATED TO REVERSIBLE LOGIC

In this section, we show the fundamentals of reversible logic, which are relevant to this manuscript.

**Definition 2.1:** A Reversible gate (denoted by n x n) is an n-input, n- output circuit which generates a unique output vector from each possible input vector [11, 12].

**Definition 2.2 :** A logic gate L is reversible if, for any output y, there is a unique input x such that, L(x)=y

If a gate L is reversible, there is an inverse gate L' which maps y to x for which L'(y)=x. It is an n-input n-output logic function in which there is a one-to-one correspondence between the inputs and the outputs means in no information loss. Because of this bijective mapping the input and output vector and can be uniquely determined from the output vector. This prevents the loss of information which is the root cause of power dissipation in irreversible logic circuits [6, 9, 12]. Cascading of reversible and inverse reversible gate is shown in Figure 1.

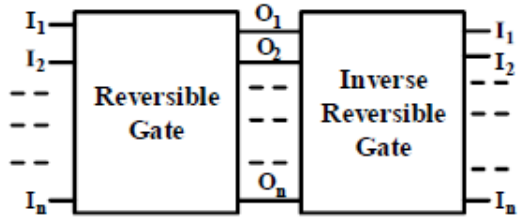

Figure 1. Cascading of reversible and inverse reversible gate.

**Definition 2.3 :** Quantum cost is one important metric, which is the number of 2x2 XOR, controlled-V or controlled-$V^+$ constitute the quantum cost of the structure. Figs 2a, b depicts the controlled-V and Controlled-$V^+$ gate [17, 19].

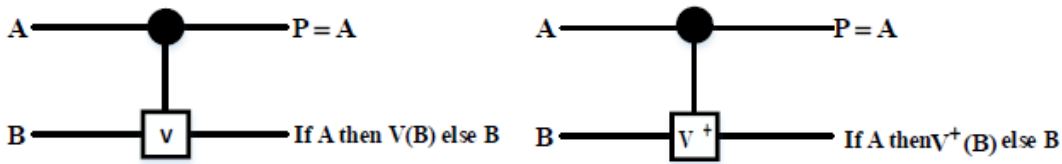

Figure 2. Quantum realization of controlled (V and $V^+$ gate)

Table1a. Functionality table of some reversible gates.

| Reversible gates | Operation Realization | Quantum cost | Quantum circuit |
|---|---|---|---|
| TG | NOT, OR, AND | 5 | |
| PG | NOT, OR, AND, XOR | 4 | |

# 3. RELATED WORKS

In this section, we implemented basic reversible gates like CNOT, TG, F2G, PG, FRG, BJN and URG in MOS transistor by using Gate diffusion input (GDI) technique. This technique reduces the power, delay and area [20] because the reversible logic circuit design that conserve information by one-to-one mapping between input and output logic.

Table 1b. Reversible 2x2 type gates.

| Block Diagram of RGate | Logic Function | MOS transistor implementation |
|---|---|---|
| 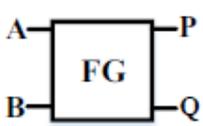 | P=A<br>Q = A ⊕ B | 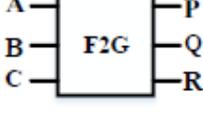 |
| 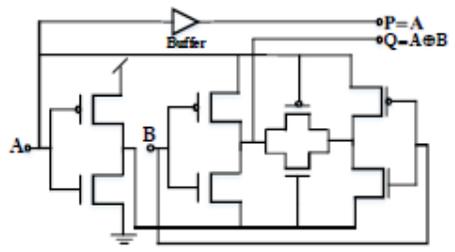 | P=A<br>Q=B<br>R = AB⊕C | 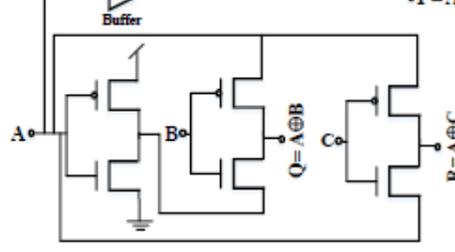 |
| 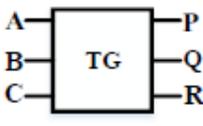 | P=A<br>Q = A ⊕ B<br>R = A ⊕ C | 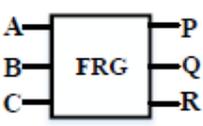 |
| 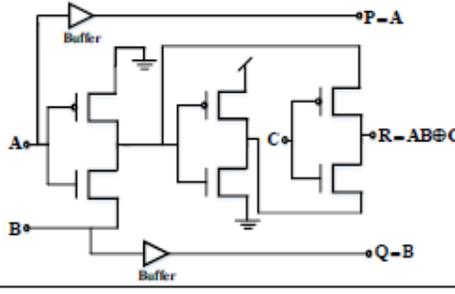 | P=A<br>Q = $\overline{A}$B + AC<br>R = AB + $\overline{A}$C | 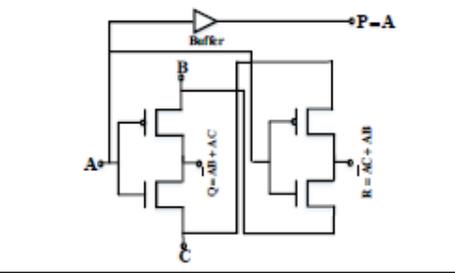 |

| Block Diagram of RGate | Logic Function | MOS transistor implementation |
|---|---|---|
| 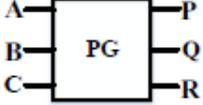 | P=A<br>Q = A ⊕ B<br>R = AB ⊕ C | 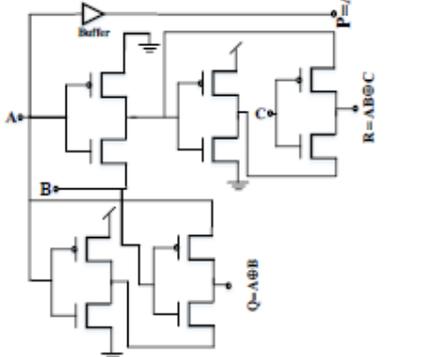 |
| 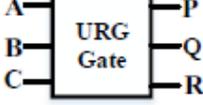 | P = (A + B) ⊕ C<br>Q = B<br>R = AB ⊕ C | 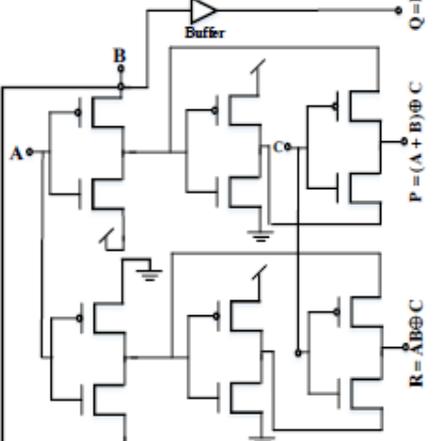 |
| 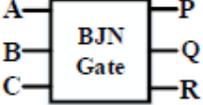 | P=A<br>Q = B<br>R = (A + B) ⊕ C | 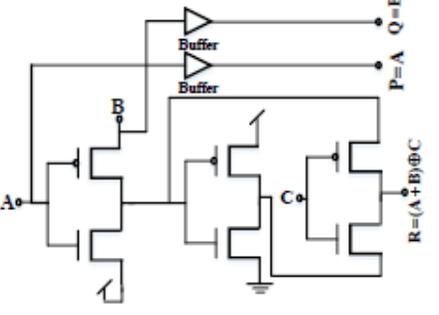 |

# 4. PROPOSED COMPRESSOR DESIGN

In this section, we design a compact reversible compressor. A compressor reduces the impact of carry (Originated from full adder) of arithmetic frame operations. To construct a compressor we using inventive0 gate for optimizing the compressor circuit. Inventive0 gate is 4x4 type reversible gate (Figure 3). The corresponding truth table (Table 3) of inventive0 gate verified that bijective mapping exist between input and output. The hardware complexity and quantum cost of inventive0 gate are 7α+4β+3δ and 10 respectively.

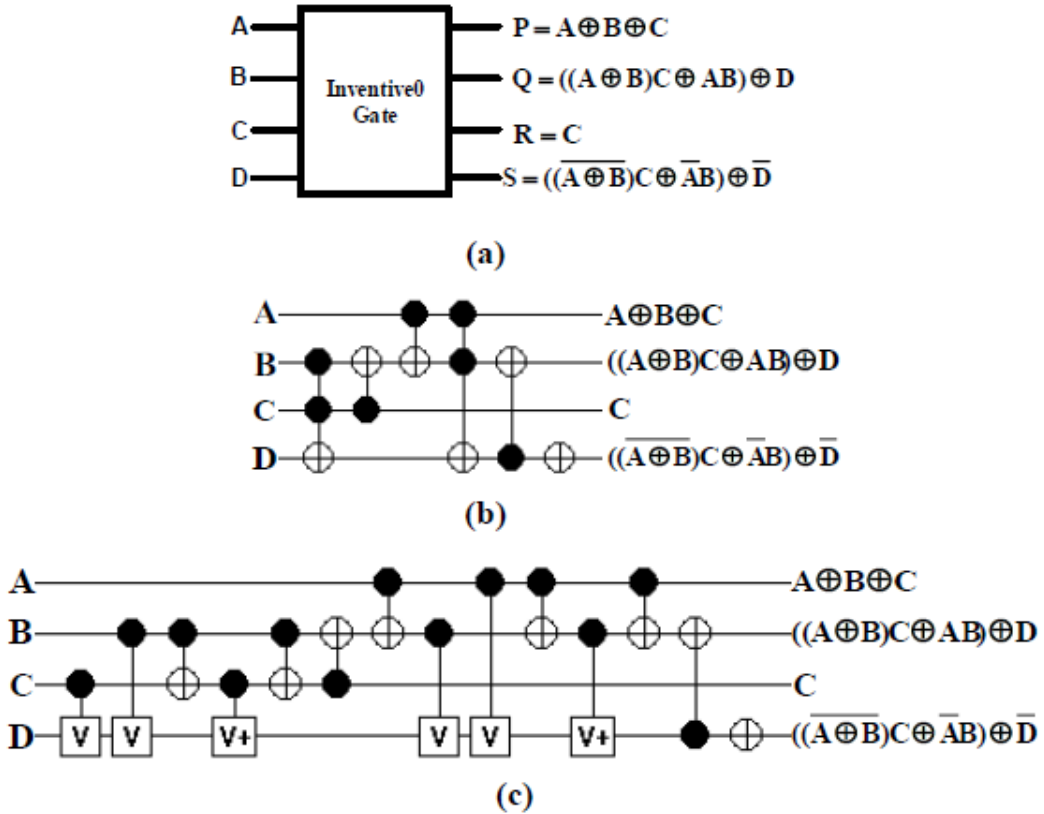

Figure 3. Reversible 4x4 Inventive0 gate. (a) Block diagram.
(b) Toffoli gate representation of Inventive0 gate.
(c) Quantum implementation of Inventive0 gate using primitive gates.

## 4.1 Some Utility of New Type of Gate

The proposed new gate, a universal gate in the sense it implements all Boolean functions. When set C=D=0, the inventive0 gate simultaneously realised the XOR logic function and the AND logic function as drawn in Fig 4a. Similarly, set C=1 and D=0 the inventive0 gate realised the XNOR logic function and OR logic function as drawn in Fig 4b. The inventive0 gate also work as a reversible full adder and full subtraction logic function, as drawn in Fig. 4c.

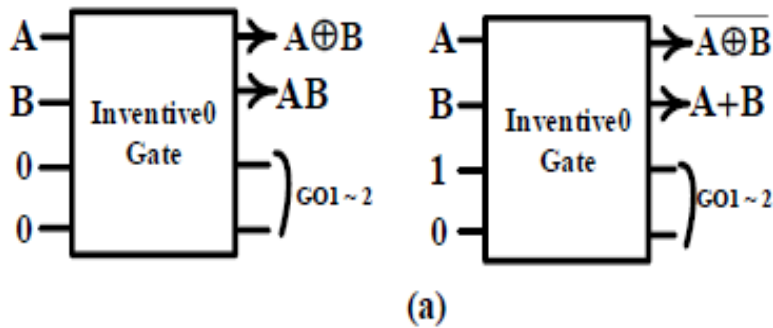

(a)

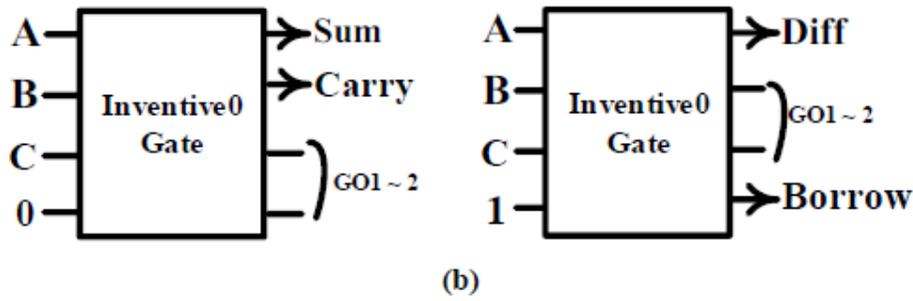

**(b)**

Figure 4. (a) Inventive0 gate utilized as XOR and AND gates.
(b) Inventive0 gate utilized as XNOR and OR gates.
(c) Inventive0 gate utilized as full adder and full subtraction

Table 3. Truth table of the 4x4 reversible Inventiv0 gate.

| Inputs | | | | Outputs | | | |
|---|---|---|---|---|---|---|---|
| **A** | **B** | **C** | **D** | **P** | **Q** | **R** | **S** |
| 0 | 0 | 0 | 0 | 0 | 0 | 0 | 1 |
| 0 | 0 | 0 | 1 | 0 | 1 | 0 | 0 |
| 0 | 0 | 1 | 0 | 1 | 0 | 1 | 0 |
| 0 | 0 | 1 | 1 | 1 | 1 | 1 | 1 |
| 0 | 1 | 0 | 0 | 1 | 0 | 0 | 0 |
| 0 | 1 | 0 | 1 | 1 | 1 | 0 | 1 |
| 0 | 1 | 1 | 0 | 0 | 1 | 1 | 0 |
| 0 | 1 | 1 | 1 | 0 | 0 | 1 | 1 |
| 1 | 0 | 0 | 0 | 1 | 0 | 0 | 1 |
| 1 | 0 | 0 | 1 | 1 | 1 | 0 | 0 |
| 1 | 0 | 1 | 0 | 0 | 1 | 1 | 1 |
| 1 | 0 | 1 | 1 | 0 | 0 | 1 | 0 |
| 1 | 1 | 0 | 0 | 0 | 1 | 0 | 1 |
| 1 | 1 | 0 | 1 | 0 | 0 | 0 | 0 |
| 1 | 1 | 1 | 0 | 1 | 1 | 1 | 0 |
| 1 | 1 | 1 | 1 | 1 | 0 | 1 | 1 |

## 4.2 Proposed reversible 4:2 Compressor circuit

In this subsection, we design a 4:2 compressor circuit using inventiv0 gate. This compressor circuit contains 2 gates (2 x inventive0), it requires four inputs (I1, I2, I3, I4) and two outputs (C2, S2) along with a carry-in (Cin) and a carry-out (C1) and two constant inputs. The proposed compressor block diagram and quantum realization as depicted in Figs 6a, b. The quantum cost of this 4:2 compressor circuit is 20, because the QC (4:2 compressor) = 2x QC (Inventive0 gate) = 2x10=20.

Total logical calculation (T) of the compressor circuit is

$$T_{inventive0} = (3\alpha) \text{ (for Q expression)} + (2\alpha+2\beta) \text{ (for R expression)} + (2\alpha+2\beta+3\delta) \text{ (for S expression)}$$
$$= 7\alpha+4\beta+3\delta$$
$$T \text{ (Design 4:2 compressor)} = 2x \ T_{inventiv0} = 2x \ (7\alpha+4\beta+3\delta) = 14\alpha+8\beta+6\delta$$

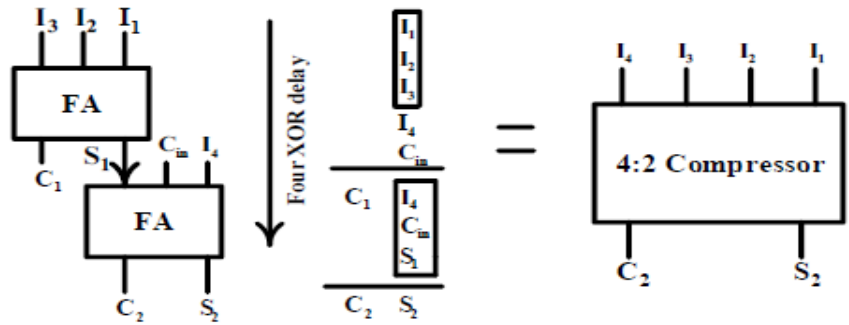

Figure 5. Conventional compressor structure.

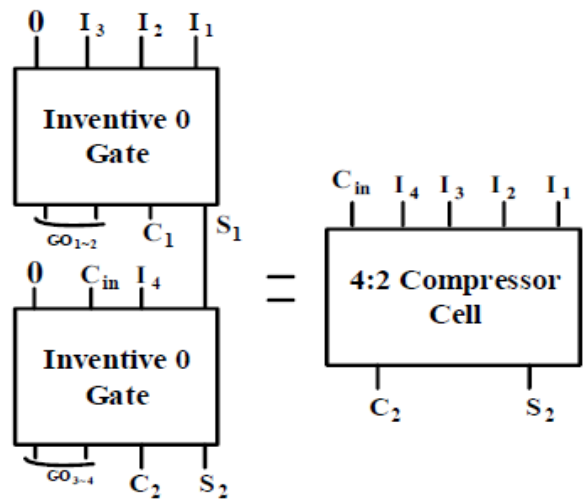

(a)

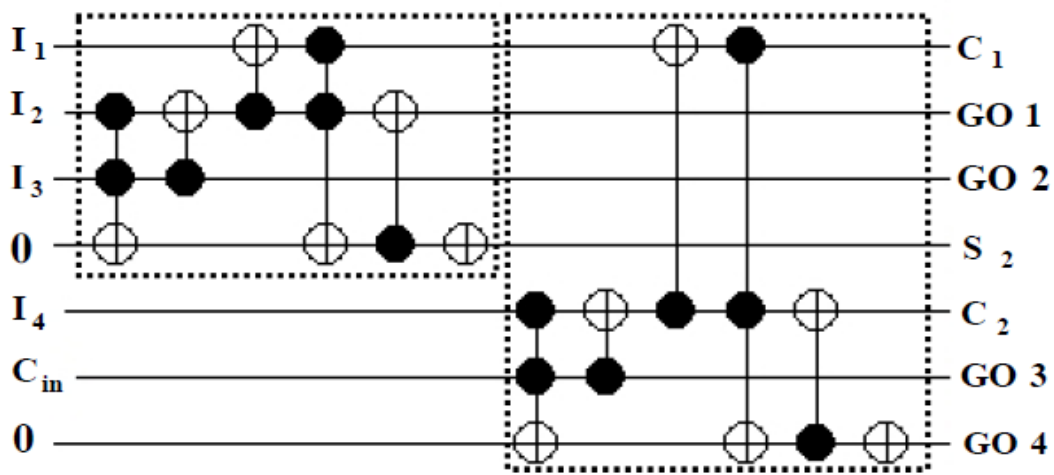

(b)

Figure 6. Proposed design of 4:2 reversible compressor.
(a) Block diagram representation.
(b) The proposed Quantum realization.

---

**Algorithm I:  Construction of reversible 4:2 compressor.**

**1. Begin**

Pick up first Inventive0 gate and pick input $I_1$, $I_2$ and $I_3$, Inputs and outputs block are considered to be levels I and O.

**2.** $I[I_1]= C[O_1]= I_1$,
$\quad I[I_2]=C[O_2]= I_2$,
$\quad I[I_3]=C[O_3]= I_3$ ;                                    // Input of $1^{st}$ Inventive0 gate.

**3.** $O[O_1]=C[O_1] \oplus C[O_2] \oplus C[O_3]$,               // Output of $1^{st}$ Inventive0 gate.
$\quad O[O_2]= ((C[O_1] \oplus C[O_2]) \, C[O_3])+ (C[O_1] \, C[O_2]) \oplus 0)$,
$\quad O[O_3]= C[O_3]= I_3$ ,
$\quad O[O_4]= (((C[O_1] \oplus C[O_2]) \oplus 1)C[O_3]+ ((C[O_1] \oplus 1) \, C[O_2]) \oplus 1)$,

**4.** Pick up second Inventive0 gate and pick input $S_1$, $I_4$ and $C_{in}$, level P

**5.** $P[I_1]= P[O_1]= S_1$,
$\quad P[I_2]=P[O_2]= I_4$,
$\quad P[I_3]=P[O_3]= C_{in}$ ;                              // Input of $2^{nd}$ Inventive0 gate.

**6.** $P[O_1]=P[O_1] \oplus P[O_2] \oplus P[O_3]$,               // Output of $2^{nd}$ Inventive0 gate.
$\quad P[O_2]=((P[O_1] \oplus P[O_2])P[O_3]+(P[O_1]P[O_2]) \oplus 0)$,
$\quad P[O_3]= P[O_3]= C_{in}$,
$\quad P[O_4]= (((P[O_1] \oplus P[O_2]) \oplus 1)P[O_3])+ \{(P[O_1] \oplus 1) \, P[O_2]\} \oplus 1)$

**End**

---

## 4.3. Proposed reversible 5:2 Compressor circuit

In this subsection, we design a 5:2 compressor circuit. This design has five primary inputs ($I_1$, $I_2$, $I_3$, $I_4$, $I_5$) and two primary outputs (C2, S2) along with two carry-in (Cin1, Cin2) and three constant inputs. The block diagram of 5:2 compressor and quantum realization as depicted in Figs 7a, b.

The quantum cost of this 5:2 compressor is 30, because the QC (5:2 compressor) = 3x QC (Inventive0gate) = 3x10=30.

T (Design 5:2 compressor) = 3 Tinventiv0= 3x $(7\alpha+4\beta+3\delta)$ = $21\alpha+12\beta+9\delta$

Total logic calculation (Circuit cost): One of the important parameter of circuit is its architectural complexity.

Where $\alpha$= Count of two input XOR gate.

$\beta$=Count of two input AND gate.

And $\delta$= Count of NOT gate.

T=Total logical calculation.

**Algorithm II:  Construction of reversible n : 2 compressor.**

 **Begin**
Here n is the reversible compressor circuit input.
**1.** If n > 4 then
**2.** Utilise one inventive0 gate as initial compressor gives the outputs of a n:2 reversible compressor.
**Loop** Step 3 for
**3.** Cascade another (n-3) inventive0 gate by using input of initial compressor and addition inputs
**4. End loop**
**5. Else**  Use initial inventive0 gate compressor
**6. End if**
**7. End**

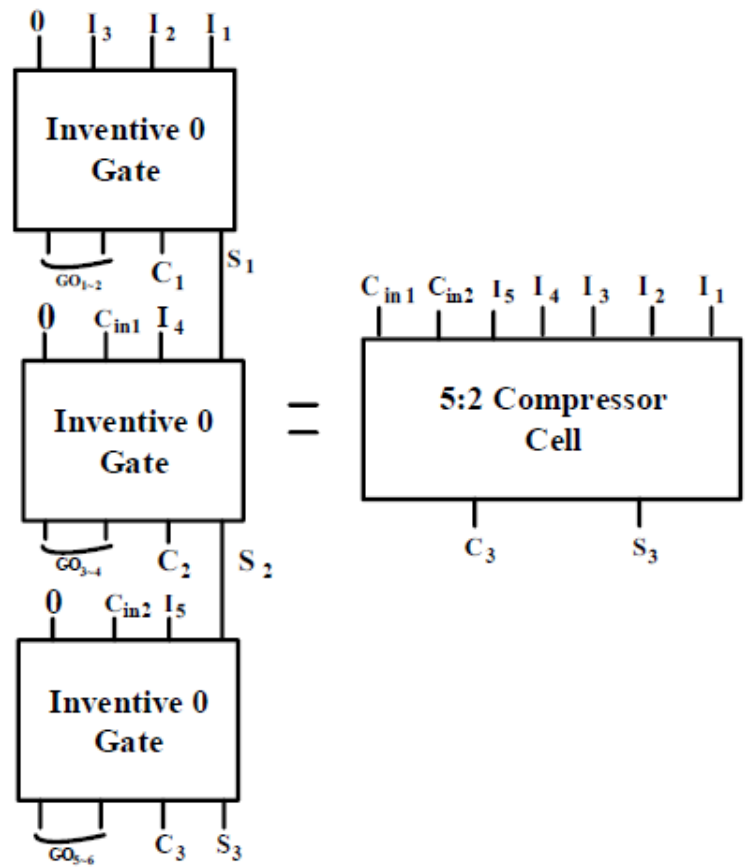

(a)

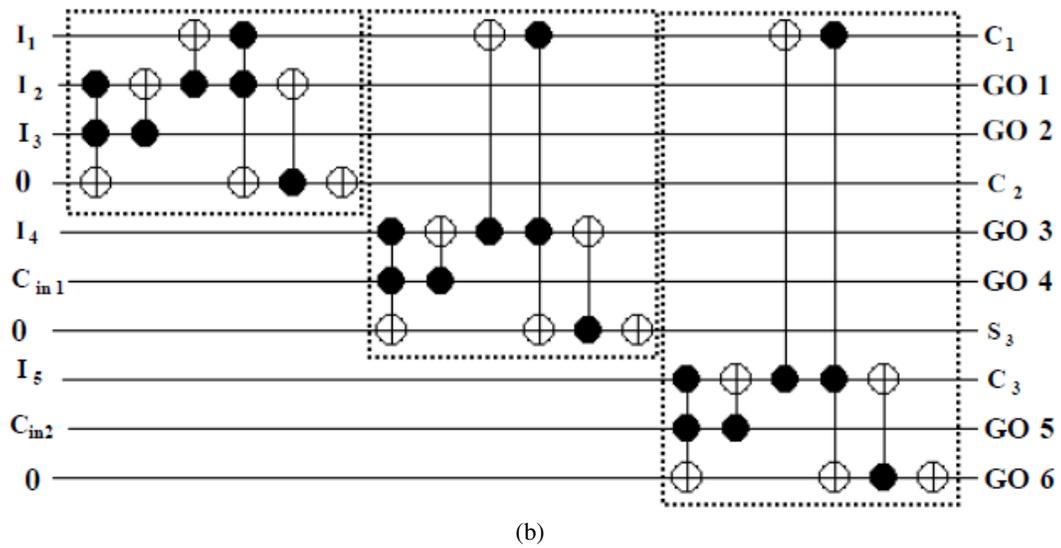

(b)

Figure 7. Proposed design of 5:2 reversible compressor.
(a) Block diagram representation.
(b) The proposed Quantum realization

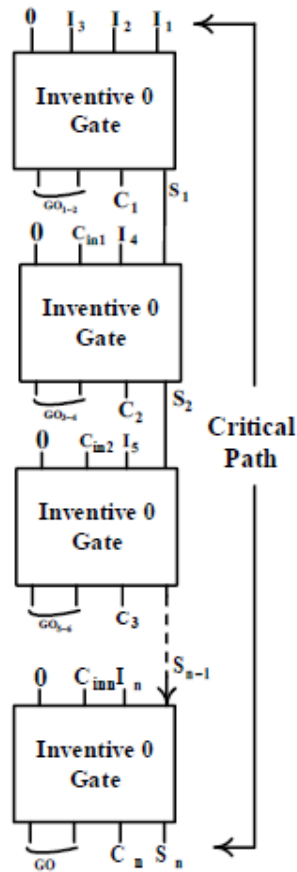

Figure 8. Design of n:2 reversible compressor

**Lemma 1:** A reversible n:2 compressor can be realised with (n-2) gates, where n is the input of compressor.

**Proof:** A 4:2 compressor consist of two inventive0 gate. Hence the total number of gates required to construct the 4:2 compressor is

NOG (4:2 compressor) = NOG (2x inventive0 gate) = 2 = 1+ (4-3)= 1+ (n-3)

Hence the statement holds for the base case n = 4

Assume that, the statement true for n = m, Hence, a reversible m:2 compressor requires at least (m-2) reversible gates.

**Lemma 2:** A reversible n:2 compressor can be realised with 10(n-3) quantum cost, where n is the input of compressor.

**Proof:** A 4:2 compressor circuit is constructed with one inventive0 gate. Hence the total quantum cost of the 4:2 compressor circuit is

QC (4:2 compressor) = QC (2x inventive0 gate) = 10 = 10 (4 - 3)

Hence the statement hold for the base case n = 4

Assume that, the statement true for n=m, Hence a reversible m:2 compressor can be realized with 10(m-3) quantum cost.

**Lemma 3:** A reversible n:2 compressor generates at least (2n-4) garbage outputs, where n is the input of compressor.

**Proof:** A 4:2 compressor circuit contains one inventive0 gate. Hence the total garbage outputs produce by 4:2 compressor is at least

GO (4:2 compressor) = GO (2x inventive0 gate) = 4= (4 x 2 - 4)

Hence, the statement holds for the base case n=4

Assume that, the statement true for n=m. Hence, a reversible m:2 compressor produce at least (2n-4) garbage outputs.

**Lemma 4:** A reversible n:2 compressor can be realised with (n-2) ancilla input, where n is the input of compressor.

**Proof:** A 4:2 compressor consist of two inventive0 gate. Hence the total ancilla input required to construct the 4:2 compressor is

CI (4:2 compressor) = CI (2x inventive0 gate) = 2 = 1+ (4-3)= 1+ (n-3)

Hence the statement holds for the base case n = 4

Assume that, the statement true for n = m, Hence, a reversible m:2 compressor requires at least (m-2) ancilla input.

## 5. PERFORMANCE EVALUATION ANALYSIS

In this section, we have shown performance evaluation analysis with the help of figures and comparative tables. We implement basic reversible gate in MOS transistor count it is depicted in Figure 8. Table 4 shows that the performance of our 4:2 compressor circuit is better compared to existing designs.

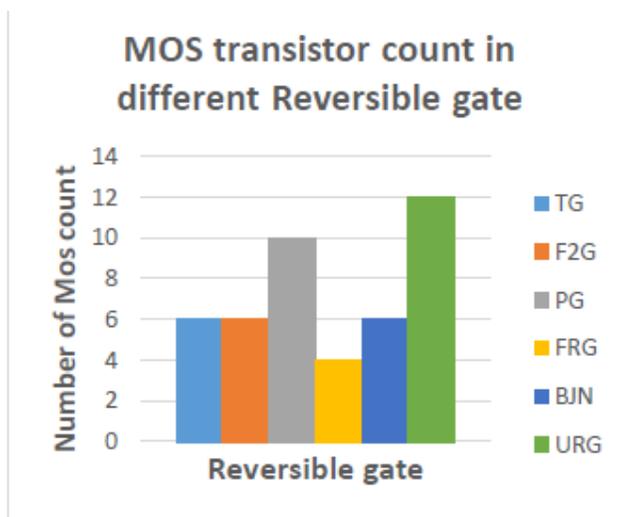

Figure 9. Number of MOS transistor count in different reversible gate

Table 4. Comparative analysis table of 4:2 compressor

| 4:2 Compressor Structures | Gate count | Constant inputs | Garbage outputs | Quantum cost |
|---|---|---|---|---|
| Proposed | 2 | 2 | 4 | 20 |
| Existing design 1 [3] | 4 | 3 | 5 | 28 |
| Existing design 2 [3] | 7 | 3 | 5 | 20 |
| Existing design 4 [3] | 2 | 2 | 3 | 26 |
| Existing design 4 [15] | 2 | 3 | 5 | 18 |

# 6. CONCLUSIONS

In this manuscript there are two main parts; the first part deals with implementing basic reversible gates CNOT, FG, TG, PG, BJN and URG gate in MOS transistor using Gate diffusion input (GDI) technique. The MOS transistor for basic reversible gate are designed by keeping in mind for the minimum MOS transistor count and also reducing the critical path. In addition, it shows a comparative performance table to the realization of the MOS transistor count. However, in the second part, we propose 4:2 compressor and 5:2 compressor circuits using inventive0 gate and also its quantum circuits. These designs are the most effective among all other existing design in terms of gate count, garbage outputs and quantum cost. Since compressor can be used in low power VLSI circuit for reducing the impact of carry (Generated from full adder operation) of arithmetic frame design, This compressor structure will be definitely used in low power circuit, ALU, quantum computers etc.